\documentclass[twocolumn,showpacs,preprintnumbers,amsmath,amssymb]{revtex4}
\usepackage{graphicx}
\usepackage{dcolumn}
\usepackage{bm}

\begin{document}


\def\KONYA{Department of Physics, University of Sel\c{c}uk, 42079 Konya,
Turkey}
\def\FIAS{Frankfurt Institute for Advanced Studies, J.W. Goethe University,
D-60438 Frankfurt am Main, Germany}
\def\MOSCOW{Institute for Nuclear Research, Russian Academy of Sciences,
117312 Moscow, Russia}
\def\GSI{GSI Helmholtzzentrum f\"ur Schwerionenforschung GmbH, D-64291 
Darmstadt, Germany}

\newcommand{\goo}{\,\raisebox{-.5ex}{$\stackrel{>}{\scriptstyle\sim}$}\,}
\newcommand{\loo}{\,\raisebox{-.5ex}{$\stackrel{<}{\scriptstyle\sim}$}\,}

\title{Theoretical study of projectile fragmentation in 
$^{112}$Sn + $^{112}$Sn and $^{124}$Sn + $^{124}$Sn reactions at 1 GeV/nucleon}

\affiliation{\KONYA}
\affiliation{\FIAS}
\affiliation{\MOSCOW}

\author{H.~Imal}    \affiliation{\KONYA}
\author{A.~Ergun}    \affiliation{\KONYA}
\author{N.~Buyukcizmeci}    \affiliation{\KONYA}
\author{R.~Ogul}   \affiliation{\KONYA}
\author{A.S.~Botvina}       \affiliation{\FIAS}\affiliation{\MOSCOW}
\author{W.~Trautmann}   \affiliation{\GSI}

\date{\today}

\begin{abstract}
We analyze the production cross sections and isotopic distributions of  
projectile-like residues in the reactions $^{112}$Sn + $^{112}$Sn 
and $^{124}$Sn + $^{124}$Sn at an incident beam energy of 1 GeV/nucleon measured
with the FRS fragment separator at the GSI laboratory. Calculations within 
the statistical multifragmentation model (SMM) for an ensemble of 
excited sources were performed with ensemble parameters determined previously
for similar reactions at 600 MeV/nucleon. The obtained good agreement with the
experiment establishes the universal properties of the excited spectator systems
produced during the dynamical stage of the reaction. It is furthermore confirmed that 
a significant reduction of the symmetry-energy term at the freeze-out stage of reduced 
density and high temperature is necessary to reproduce the experimental isotope 
distributions. A trend of decreasing symmetry energy for large neutron-rich fragments 
of low excitation energy is interpreted as a nuclear-structure effect. 

\end{abstract}


\maketitle

\section{Introduction}
The study of nuclear fragmentation and multifragmentation reactions is 
of particular interest due to both, their important practical applications 
as well as the new research opportunities that these reactions provide. The latter 
include the investigation of the equation of state of nuclear matter, 
of the composition of nuclear matter at subnuclear densities, and of
phase transitions in nuclear systems. Information of this kind can be related
to processes taking place during the collapse and explosion of massive stars 
and in the formation of neutron stars~\cite{Lattimer01,Botvina10}. 
The isospin composition of the produced fragments has been found to be 
especially important because it can be used for determining the strength of 
the symmetry energy during fragment formation in the hot and diluted 
environment~\cite{Botvina02,Ono03,LeFevre05,Ogul11},
which is crucial for weak reaction rates in stellar matter~\cite{Botvina10}. 

Mid-peripheral heavy-ion collisions at relativistic energies provide us 
with the possibility to study the production of isotopes as a result of
fragmentation and multifragmentation of the colliding nuclei. 
Recently, experiments for two symmetric systems $^{124}$Sn + $^{124}$Sn 
and $^{112}$Sn + $^{112}$Sn, both at an incident energy of 1 GeV/nucleon, 
were performed by the FRS collaboration at the GSI laboratory~\cite{Fohr11}. 
The high-resolution magnetic FRagment Separator (FRS) was used 
for the separation and identification of the reaction products. 
The initial neutron to proton ratios ($N/Z$) of the symmetric systems are 
1.24 for $^{112}$Sn and 1.48 for $^{124}$Sn. The measured isotopic cross 
sections of identified fragments from these two reactions are reported in
tabulated form in Ref.~\cite{Fohr11}.

In a previous study with the ALADIN forward-spectrometer, the fragmentation of 
stable $^{124}$Sn and radioactive $^{107}$Sn and $^{124}$La projectiles at 
600 MeV/nucleon has been measured and analyzed~\cite{Ogul11}. 
In particular, it was found that the charge and isotope 
yields, fragment multiplicities and temperatures, and correlations of various 
fragment properties can be well described within the Statistical 
Multifragmentation Model (SMM, Ref.~\cite{bondorf95}).
An ensemble of excited sources was assumed to represent the collision systems
after the initial non-equilibrium part of the reaction and chosen as starting configuration
for the statistical description of the subsequent reaction stages. Its general form
in the plane of source mass and excitation energy was adapted from previous 
studies~\cite{botvina92,botvina95,xi97} and the sensitivity of the experimental
observables to the parameters required for the liquid-drop description of the 
primary excited fragments was evaluated~\cite{Ogul11}.   
With the present work, we demonstrate that the same approach can be used to describe
the recent FRS data for similar projectiles and that comparable results concerning 
the required model parameters are obtained. With the new FRS data, the analysis can
be extended to include the distributions of heavier isotopes up to close to the
initial projectile mass. 

As shown previously, the symmetry energy represents the main model parameter 
governing the mean $N/Z$ values, the isoscaling parameters, and the 
isotopic composition of the fragment yields. For more violent collisions associated 
with larger particle and fragment multiplicities, its strength needs to be reduced 
if an adequate description
of the experimental data is to be achieved~\cite{LeFevre05,Ogul11,Souliotis07,Hudan09}.
This observation was also made in the analysis of the experimental data of Liu et al.~\cite{Liu04} 
obtained at the MSU laboratory for the same projectiles and targets that were studied
with the FRS, i.e. $^{124}$Sn + $^{124}$Sn and $^{112}$Sn + $^{112}$Sn, but at the lower energy
of 50 MeV/nucleon~\cite{Ogul09,Buyukcizmeci11,Buyukcizmeci12}.

\section{Statistical Approach To Multifragmentation}

It has been repeatedly demonstrated that the statistical 
multifragmentation model (SMM, Ref.~\cite{bondorf95}) is a useful 
tool for describing the fragment production in peripheral heavy-ion collisions 
at high energy~\cite{Ogul11,botvina95,xi97,EOS}. In the 
present work, we consider the ensemble approach with the same parameters 
that were used for the interpretation of the ALADIN data~\cite{Ogul11}. 
The general properties of the considered ensembles of residual nuclei, found to 
be quite adequate for describing the multifragmentation of relativistic projectiles, 
are given in Ref.~\cite{botvina95}. 
The excited residues form a broad distribution in the energy vs. mass plane, extending from 
large masses near 
the projectile mass and low excitation energies up to sources of small mass but high excitation energy 
in the vicinity of the nucleon binding energies. The $N/Z$ ratios of the sources are taken to be 
those of the initial projectiles.

\begin{figure} [tbh]
\begin{center}
\includegraphics[width=8.6cm,height=11cm]{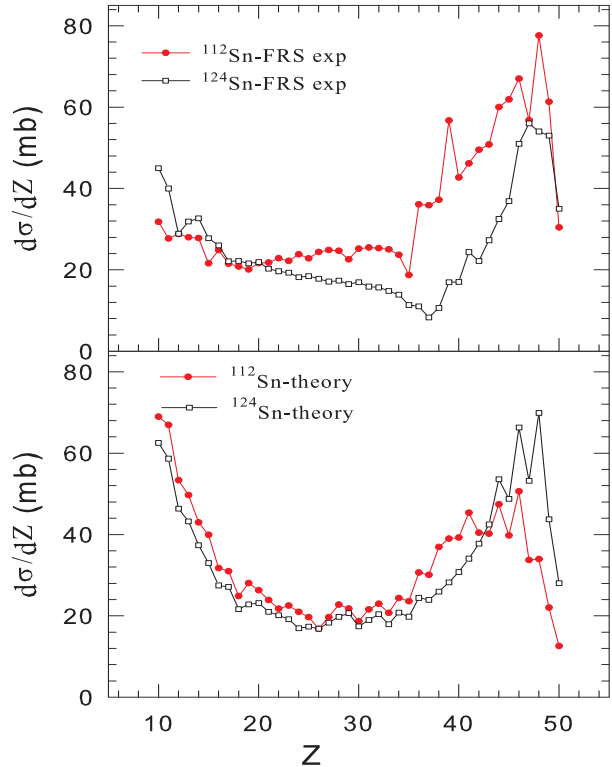}
\end{center}
\caption{\small{(color online) Top panel: production cross sections of the measured 
projectile fragments as a function of the fragment atomic number $Z$ for the two reactions
(from Ref.~\protect\cite{Fohr11}). Bottom panel: 
the corresponding distributions as obtained from ensemble calculations with standard 
parameters (for details see text in Section II).}}  
\end{figure}

In the SMM, it is assumed that a statistical equilibrium is reached within
a low-density freeze-out zone. 
The breakup channels are composed of nucleons and nuclear fragments, and 
the conservation laws (energy, momentum, angular momentum, mass number $A$ 
and atomic number $Z$) are taken into consideration. Besides the breakup 
channels, the compound-nucleus channels are also included, and the competition 
between all channels is permitted. In this way, the SMM covers the 
conventional evaporation and fission processes occurring at low excitation 
energy as well as the transition region between the low and high energy 
de-excitation regimes. In the thermodynamic limit, the SMM is consistent 
with a liquid-gas type phase transition in which the liquid phase is represented 
by an infinite nuclear cluster \cite{Das98}, permitting the connection with 
the astrophysical case~\cite{nihal13}. 

For finite nuclear systems, the SMM version developed in 
Refs.~\cite{bondorf95,botvina85,botvina87} is used. 
It represents the main version used previously for successful comparisons with 
a variety of experimental data 
\cite{Ogul11,botvina95,xi97,EOS,dagostino,ganil,FASA}. We calculate the 
contributions of all breakup channels partitioning the system into various 
species. The decay channels are generated by a Monte Carlo method according 
to their statistical weights. Light fragments with mass number $A\leq4$ 
and charge number $Z\leq2$ are considered as elementary particles (nuclear 
gas) with their corresponding spins and translational degrees of freedom. 
The fragments with mass number $A>4$ are treated as heated nuclear 
liquid-drops. In this way one can study the nuclear liquid-gas 
coexistence in the freeze-out volume. The free energies $F_{A,Z}$ of  
fragments are parameterized as the sums of the bulk, surface, Coulomb and 
symmetry energy contributions
\begin{equation}
F_{A,Z}=F_{A,Z}^B+F_{A,Z}^S+E_{A,Z}^C+E_{A,Z}^{sym}.
\end{equation}
The bulk contribution is given by $F_{A,Z}^B=(-W_0-T^2/
\varepsilon_0)A$, where $T$ is the temperature, the parameter $
\varepsilon_0$ is related to the level density, and $W_0=16$ MeV
is the binding energy of infinite nuclear matter. 
The contribution of the surface energy is 
$F_{A,Z}^S=B_0 A^{2/3}[(T_{\rm c}^2-T^2)/(T_{\rm c}^2+T^2)]^{5/4}$, 
where $B_0=18$ MeV is the surface energy term, 
and $T_{\rm c}=18$ MeV the critical 
temperature of infinite nuclear matter. The Coulomb  
contribution is $E_{A,Z}^C=cZ^2/A^{1/3}$, where c denotes the 
Coulomb parameter obtained in the Wigner-Seitz approximation, 
$c=(3/5)(e^2/r_0)(1-( \rho / \rho_0)^{1/3})$, with the charge unit 
e, $r_0=1.17$ fm, and the normal nuclear-matter density $\rho_0 = 0.15$~fm$^{-3}$. 
And finally, the symmetry term is 
$E_{A,Z}^{sym}= \gamma (A-2Z)^2/A$, where $\gamma =25$ MeV is the 
symmetry energy parameter. All the parameters given above are 
taken from the Bethe-Weizs\"acker formula and correspond to the 
assumption of isolated fragments with normal density 
unless their modifications in the hot and dense freeze-out configuration 
follow from the analysis of experimental data. For the freeze-out density, 
one-third of the normal nuclear matter density is assumed, as in previous 
studies and qualitatively consistent with independent experimental 
determinations~\cite{Fritz99,Viola04}.

\begin{figure} [tbh]
\begin{center}
\includegraphics[width=8.5cm]{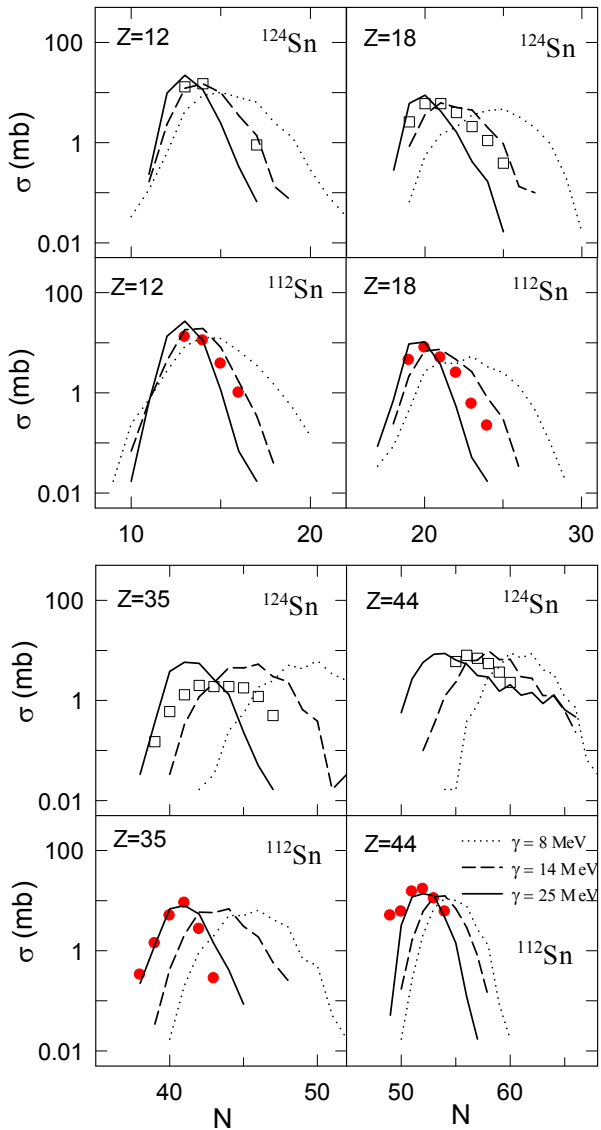}
\end{center}
\caption{\small{(color online) Predicted (lines) and measured (symbols,
from Ref.~\protect\cite{Fohr11}) isotope distributions for final 
fragments of the projectiles $^{124}$Sn and $^{112}$Sn with atomic numbers $Z=12$, 18, 35, 
and 44. Calculations are shown for three different choices of the symmetry-term coefficient
$\gamma=8$ (dotted), 14 (dashed), and 25 MeV (full lines).}} 
\end{figure}
 
After formation in the freeze-out volume the hot fragments undergo  
secondary de-excitation and propagate in the mutual Coulomb field. 
Their secondary decay includes evaporation, fission, and Fermi-break-up 
processes. The corresponding models are similar to those used for the
description of low energy reactions \cite{bondorf95,botvina87,eren13}; 
they do account, however, for modifications of fragment properties 
under freeze-out conditions~\cite{henzlova10,Buyukcizmeci05}. 

Especially, the modification of the fragment symmetry energy must be taken into account in the first 
de-excitation steps, as the hot fragments are still surrounded by other species. At the end of the 
evaporation cascade, the standard properties    
will have to be restored as explicitly described in Ref.~\cite{Buyukcizmeci05}. In the  
calculations, this is realized with a linear interpolation between these two limiting cases in the interval 
of excitation energies below $E_x^{\rm int} = 1$~MeV/nucleon. 
In this interval, the mass of a nucleus with mass and atomic numbers $A$ and $Z$ evolves with the excitation 
energy $E_x$ as
\begin{equation}
m_{A,Z}=m_{ld}(\gamma) \cdot x + m_{st} \cdot (1-x)
\end{equation}
where $x=E_x/A/E_x^{\rm int}$ and $x\leq$1. The excitation energy $E_x$ is determined from the energy 
balance, taking into account the mass $m_{A,Z}$ at the given excitation. 
The liquid-drop mass $m_{ld}$
is that of hot fragments as adopted in the SMM (without temperature and density dependences),
\begin{eqnarray}
m_{ld}(\gamma)=m_{n}(A-Z)+m_{p}Z-AW_0+B_0A^{2/3} \nonumber\\
+\gamma\frac{(A-2Z)^2}{A}+\frac{3e^2Z^2}{5r_0 A^{1/3}},
\end{eqnarray}
where $m_{n}$ and $m_{p}$ are the masses of free neutrons and protons and $W_0$ and $B_0$ the volume and 
surface term coefficients, respectively. 

The standard masses $m_{st}$ are taken from 
the nuclear mass tables or, if experimental masses are not available, the mass formula of Myers and 
Swiatecki~\cite{ms} is used. 
It has also been confirmed that, with the standard symmetry-term coefficient $\gamma=25$~MeV, 
the interpolation procedure leads to very similar results as a standard evaporation that uses the 
ground-state masses $m_{st}$ throughout.

\section{Charge and Isotopic Distributions}

Cross sections for projectile fragmention in the two studied reactions 
$^{124}$Sn + $^{124}$Sn and $^{112}$Sn + $^{112}$Sn at 1 GeV/nucleon are
shown in Fig.~1 as a function of the atomic number $Z$ of the final fragments.
The experimental results (top panel) are obtained by summing up the isotope 
yields given in the tables of Ref.~\cite{Fohr11}. 
The bottom panel shows the theoretical results obtained from ensemble calculations 
performed for 500 000 reaction events with the standard SMM parameters given in 
Section II and with the ensemble
parameters used in the analysis of the ALADIN data in Ref.~\cite{Ogul11}. 
To permit a more quantitative comparison of the model results with the experimental 
data, the SMM ensemble calculations were globally normalized with respect to the 
measured elemental cross sections in the interval $20\leq Z \leq 25$. 
The obtained factors are 0.00334 mb and 0.00344 mb per theoretical 
event for $^{124}$Sn and $^{112}$Sn projectiles, respectively. 
The observed agreement between the experimental and theoretical results is, at most,
qualitative and considerable differences exist. For $^{124}$Sn in particular, the
measured yields in the range $30\leq Z \leq 45$ seem low, with respect to the calculations
as well as relative to the experimental yields for $^{112}$Sn. To a large extent, this
is due to the fact that the isotope distributions were not always fully covered in the 
experiment, thus causing the observed distortions of the $Z$ distributions from their 
known general form in high-energy reactions 
(see, e.g., Refs.~\cite{bondorf95,botvina90,scheidenberger04}). 

Cross section differences are also observed for individual isotopes from different 
experiments. For example, 
F\"{o}hr et al.~\cite{Fohr11} report a cross section of $19 \pm 4$~mb 
for the production of $^{22}$Ne in the fragmentation of $^{124}$Sn projectiles at 
1 GeV/nucleon while the corresponding cross section reported by Ogul et al.~\cite{Ogul11}
for the same reaction at 600 MeV/nucleon is approximately 40~mb. A decrease with energy
is not expected but it has to be considered that the solid-angle acceptance of the 
FRS is much smaller than that of the ALADIN spectrometer and that, when evaluating 
the total production cross sections for intermediate mass fragments (IMF), 
the assumption of an isotropic emission 
in the rest frame of the considered species was adopted in Ref.~\cite{Fohr11}. 
This leads to an underestimation if the transverse momentum distributions are wider than
the longitudinal distributions that are actually measured with the FRS.
Anisotropies of this kind may be caused by a side-motion of the excited projectile 
residues after the dynamical reaction stage that is especially seen in the 
multifragmentation regime~\cite{urbon80,viola99}. 
The velocity characteristics of fragments previously investigated with the FRS 
show the effects of entering into the multifragmentation 
regime~\cite{napolitani04}, and the problem of underestimating the yields of fragments 
with Z$\loo$14 is admitted in Ref.~\cite{Fohr11}. For these reasons, 
less weight is given to the absolute IMF yields in the present analysis which, 
in the following, will be focussed on the relative isotope distributions. 

\begin{figure} [tbh]
\begin{center}
\includegraphics[width=8.6cm,height=11cm]{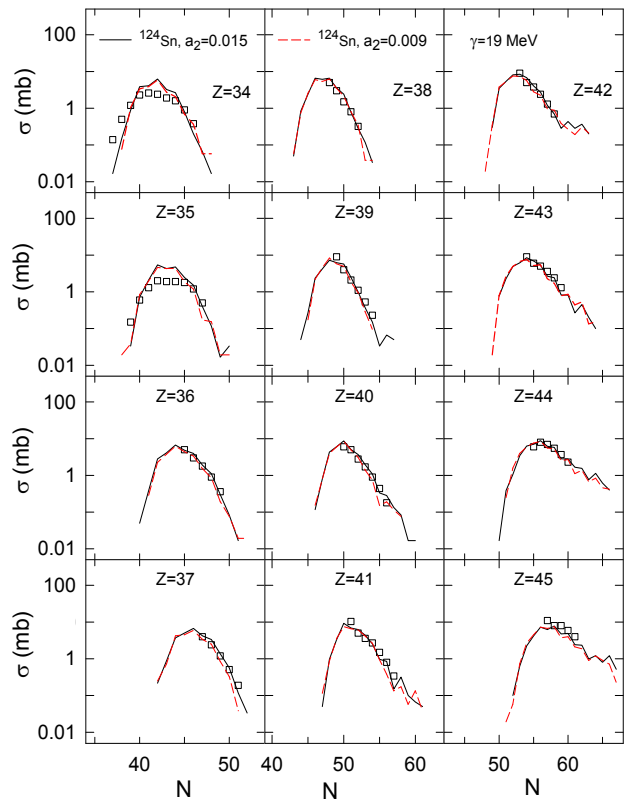}
\end{center}
\vskip -0.2cm
\caption{\small{(color online) Predicted isotopic cross-sections for final 
fragments with atomic numbers $34 \le Z \le 45$ from the fragmentation of $^{124}$Sn 
projectiles for two ensemble parameters $a_2 = 0.015$~MeV$^{-2}$ (full line) and 
$a_2 = 0.009$~MeV$^{-2}$ (dashed)
in comparison with the experimental data from Ref.~\protect\cite{Fohr11} (open squares).
The symmetry-term coefficient $\gamma = 19$~MeV was used.}} 
\end{figure}

The parameter dependence of the predicted fragment yields was investigated in detail previously~\cite{Ogul11}. 
Both, the parameterization of the ensemble of hot sources as well as the description of the produced fragments in the 
hot environment were included in this study. It was found that 
possible modifications of the symmetry-energy term appear rather exclusively in the width and positions of the fragment 
isotopic distributions while other observables as, e.g., fragment multiplicities or $Z$ distributions, are not 
particularly affected. For the present case, this is illustrated
in Fig. 2 which shows the variation of the calculated isotopic distributions with 
the symmetry-energy coefficient $\gamma$ for the final fragments with $Z=12$, 18, 35 and 44.
The neutron numbers of the strongest isotope of an element and the widths of the distributions
vary strongly with the $\gamma$ parameter. A reduced $\gamma$ causes the isotope distributions
to be wider. It also leads to a larger neutron content of the final fragments since the 
probability for charged-particle emission during the secondary de-excitation of neutron-rich 
primary fragments becomes higher~\cite{Ogul11,henzlova10,Buyukcizmeci05}. 
The measured yields are best reproduced with $\gamma = 14$~MeV in the case of the lightest 
fragments with $Z=12$ while the standard $\gamma = 25$~MeV seems more appropriate for $Z=44$. 

Comparing the $^{124}$Sn and $^{112}$Sn fragmentations, the measured distributions are very 
similar for the lighter fragments ($Z=12, 18$), shifted with respect to each other by 
approximately one mass unit, but qualitatively different for the heavier species which 
are not produced in multifragmentation. 
The mass distributions of the $Z=35$ and 44 fragments are relatively narrow in good agreement
with the predictions for $\gamma = 25$~MeV in the case of the $^{112}$Sn fragmentation but
much wider with a tail to larger mass numbers for $^{124}$Sn. This had already been noticed 
by the authors of Ref.~\cite{Fohr11} and was explained by showing the overall good agreement
of the yield distributions 
with the predictions obtained from the empirical EPAX parametrization~\cite{Suemmerer00}.
As shown by Charity~\cite{Charity98}, the EPAX yield distributions approach the 
evaporation-attractor-line (EAL) if the excitation energies of the primary residues 
are sufficiently high, so that neutron and charged-particle evaporations will effectively
compete. In the present case, the EAL is more easily reached from
the neutron-poor $^{112}$Sn residues and, in fact, the most probable isotopes are found to lie close to 
this line as obtained with GEMINI evaporation calculations by Charity~\cite{Charity98} or, 
equivalently, also with the secondary-evaporation calculations implemented in the 
SMM. The lighter fragments from $^{124}$Sn ($Z = 12$ and 18) 
are also very close but a memory effect is clearly observed, 
reminiscent of the neutron richness of the initial sources.

For neutron-rich nuclei with lower excitations, neutron evaporation dominates and the mass 
number of the final product nucleus will be mainly determined by its initial excitation energy.
While this is valid for individual source nuclei it does not affect much the final isotope 
distributions resulting from the ensemble of excited sources studied here. This has been confirmed 
with test calculations using varying ensemble parameters which show that the variations of the 
de-excitation chains are compensated by the change of the distribution of source nuclei populating 
a particular element. An example is given in Fig. 3 which presents the
results for products with $34 \le Z \le 45$ from $^{124}$Sn fragmentation for ensembles
generated with parameters $a_2 = 0.015$ (the standard value) and 0.009~MeV$^{-2}$.    
This parameter determines the curvature in the expression
\begin{equation} 
A_s/A_0 =1-a_1(E_x/A_s)-a_2(E_x/A_s)^2
\end{equation} 
describing the average mass number $A_s$ of the equilibrated 
sources as a function of their excitation energy $E_x$ (in MeV). Here $A_0$ is the projectile mass and
the parameter $a_1$ is taken as $a_1=0.001$~MeV$^{-1}$. 
The smaller value for $a_2$ produces an ensemble with higher excitation energy, amounting to an 
increase $\Delta E_x/A = 0.4$ to 0.8~MeV 
for source nuclei with mass numbers larger than about 90\% of the 
projectile mass (cf. Fig.~5 in Ref.~\cite{Ogul11}). The resulting modifications of the
isotope distributions are, apparently, very small. 
As a consequence, also the isotope distributions for heavier fragments from the $^{124}$Sn 
fragmentation, peaking several mass units away from the EAL ($Z=35$ and 44 shown in Fig. 2),
retain their sensitivity to the strength of the symmetry energy. 

\begin{figure} [tbh]
\begin{center}
\includegraphics[width=8.6cm,height=11cm]{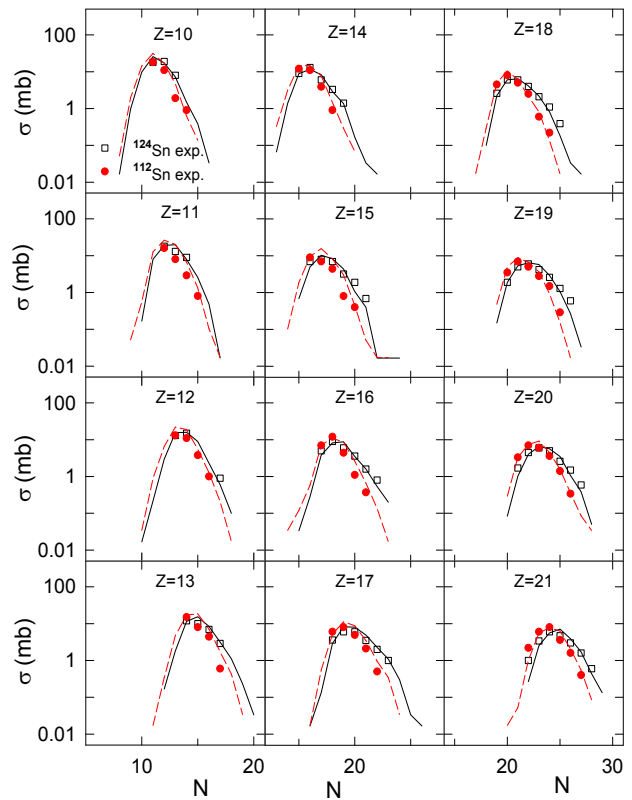}
\end{center}
\vskip -0.2cm
\caption{\small{(color online) Predicted (lines) and measured (symbols, 
from Ref.~\protect\cite{Fohr11}) isotopic cross-sections for final 
fragments with atomic numbers $10 \le Z \le 21$ from the fragmentation of $^{124}$Sn 
(open squares and full lines) and $^{112}$Sn projectiles (closed circles and dashed lines).
The symmetry-term coefficients $\gamma$ used in the calculations are given in Table I.}} 
\end{figure}

\begin{figure} [tbh]
\begin{center}
\includegraphics[width=8.6cm,height=11cm]{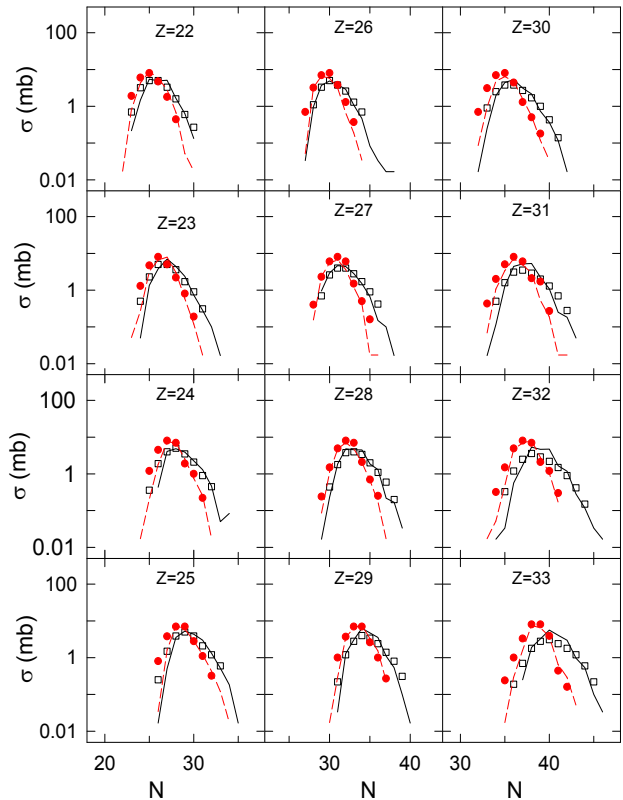}
\end{center}
\vskip -0.2cm
\caption{\small{(color online) As Fig. 4 but for the final fragments with 
atomic numbers $22 \le Z \le 33$.}}
\end{figure}

The evolution of the final isotope distributions with atomic numbers from $Z=10$ to $Z=45$
is shown in Figs. 4 - 6 in a direct comparison of the neutron-rich and neutron-poor cases.
The corresponding predictions had been calculated with standard parameters as, e.g., 
$a_2 = 0.015$~MeV$^{-2}$ and $B_0 = 18$~MeV and with various symmetry-energy coefficients
$\gamma$. In the figures, they are shown for the $\gamma$ values found to be most adequate
for the selected five groups of elements listed in Table~I. As expected 
from Fig. 2, the favored symmetry-term coefficient increases from $\gamma = 16$~MeV for 
$Z=10-17$ to $\gamma = 25$~MeV for $Z=38-45$ and $^{112}$Sn. In the case of $^{124}$Sn,
the obtained $\gamma$ values are similar for the smaller fragments but don't exceed
$\gamma = 20$~MeV as the fragment $Z$ approaches that of the projectile.

The agreement of the calculated cross sections with the measured data is, overall, very 
satisfactory. We emphasize again that the calculations are obtained with the parameter set
determined in the analysis of the ALADIN fragmentation data~\cite{Ogul11} and with a global
normalization in the cross section interval $20\leq Z \leq 25$. In particular also the yields
of heavier isotopes with atomic numbers $Z \approx 35$ to 45 are quite well reproduced (Fig.~6). 
Here, in the case of $^{124}$Sn, the isotopic distributions
were only partly covered in the experiment which explains why the integrated cross sections 
appear rather low in this part of the $Z$ spectrum (Fig.~1). It has also the effect 
that the determination of the optimum $\gamma$ value is more difficult in the neutron-rich case.
Except for $Z = 35$ and 44, the maxima of the isotopic yield distributions are not 
unambiguously determined (Fig.~6). It is only because of the larger sensitivity (cf. Fig.~2) that 
the precision is still of the order of $\Delta \gamma = \pm 1$ also for $^{124}$Sn. 

\begin{table} [tbh]
\centering
\caption{Symmetry-energy coefficient $\gamma$ chosen for the calculations of the  
isotopic yield distributions for the listed five element groups on the basis of an optimum 
reproduction of the experimental results for the two projectile cases.}
\begin{tabular}{ccccc}
\hline 
\hline
Z & &$^{112}Sn$ & &$^{124}Sn$\\
 intervals& &~~~~~~~~~~~~~~~~~$\gamma$(MeV)~~~~~~~~~~~~~~~& &$\gamma$(MeV)\\
\hline
Z=10-17 & & 16 & & 16\\
Z=18-25 & & 19 & & 18\\
Z=26-31 & & 21 & & 20\\
Z=32-37 & & 23 & & 19\\
Z=38-45 & & 25 & & 18\\
\hline
\hline 
\end{tabular}
\end{table}

\begin{figure} [tbh]
\begin{center}
\includegraphics[width=8.6cm,height=11cm]{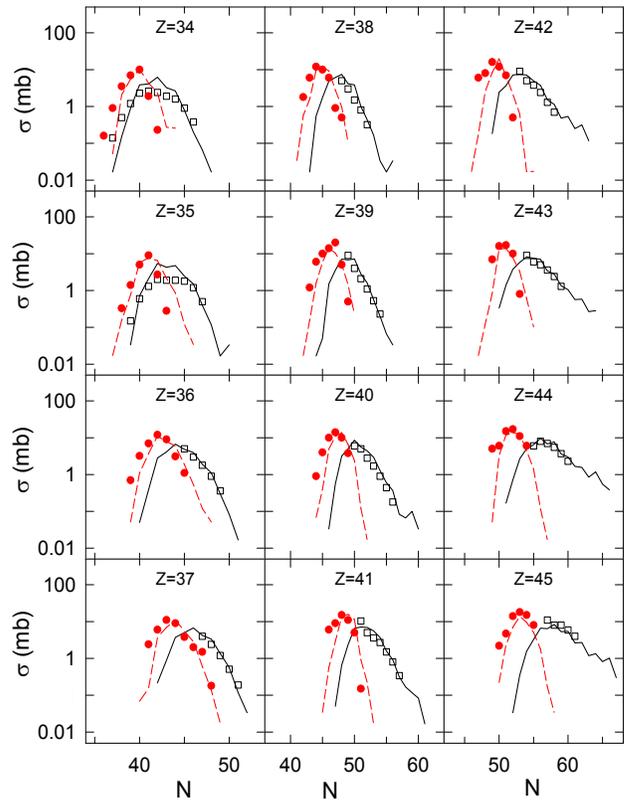}
\end{center}
\vskip -0.2cm
\caption{\small{(color online) As Fig. 4 but for the final fragments with 
atomic numbers $34 \le Z \le 45$.}}
\end{figure}

The FRS data are inclusive but smaller fragments are, nevertheless, known to be predominantly
produced in more violent collisions. The global behavior of $\gamma$ as a function of the
fragment mass is, therefore, fully consistent with the impact-parameter dependence deduced
from the exclusive ALADIN data for fragmentations at 600 MeV/nucleon. 
Studied there as a function of the bound charge $Z_{\rm bound}$ used for impact-parameter sorting,
the symmetry-term 
coefficient was found to decrease rapidly with increasing multiplicity of fragments and 
light particles from the decay of the excited spectator systems. 
Interpreted as resulting from the overall reduced density at breakup and from the hot environment
modifying fragment properties, this observation is shown here to extend into the regime of 
medium-size fragment production up to about $Z = 30$.  

A more quantitative comparison shows that the present $\gamma = 16$~MeV deduced for $Z=10-17$ 
corresponds approximately to the result for the bin of reduced bound charge $0.6<Z_{\rm bound}/Z_0 <0.8$ 
of the ALADIN data. 
Bins of smaller reduced $Z_{\rm bound}$ exhibit fragment spectra that decrease as a function of $Z$ with average 
values $<Z> \approx 10$ and below. 
The symmetry-term coefficient $\gamma$ determined for these more central bins continues to drop to considerably 
smaller values~\cite{Ogul11}.

In the same study, it was also explictly shown that the required coefficient $\gamma$ is smaller than what 
could be expected by including a surface-symmetry term into the model. Four different parametrizations were used 
to calculate an effective symmetry energy, averaged over the experimental set of partitions observed in a 
particular $Z_{\rm bound}$ bin. The smaller fragments produced at higher excitations cause indeed the effective 
mean symmetry term to decrease with decreasing $Z_{\rm bound}$  
but at a slower rate than that resulting from the analysis of the experimental data~\cite{Ogul11}. 

A similar comparison can be made for the present case. By using the volume symmetry and surface symmetry 
coefficients $a_{\rm sym}^{\rm vol}= 28.1$~MeV and $a_{\rm sym}^{\rm surf}= 33.2$~MeV of the Myers and 
Swiatecki mass formula~\cite{ms} in the expression
\begin{equation}
\gamma_{\rm eff} = a_{\rm sym}^{\rm vol} - a_{\rm sym}^{\rm surf}/A^{(1/3)}
\end{equation}
effective symmetry-term coefficients may be obtained that explicitly reflect the surface effect.
For the bin of smallest fragments studied here, $Z=10-17$, typical mass numbers are $A=30$ at the end of the 
de-excitation sequence (cf. panel $Z=14$ in Fig. 4) or masses around $A=34-38$ of the hot sources populating 
this bin (as given by the SMM calculations). The corresponding $\gamma_{\rm eff}=17.4$~MeV for $A=30$ and 
$\gamma_{\rm eff}=18.0$~MeV for $A=36$ are not much but distinctly larger than the $\gamma=16$~MeV deduced from
the data (Table I) in full agreement with the trends reported in Ref.~\cite{Ogul11}. 

The usefulness of such
a comparison has to be judged with some caution because the shell correction terms used in the
Myers and Swiatecki mass formula are ignored here and, as stated above, the standard symmetry-term 
coefficient $\gamma=25$~MeV is needed to obtain similar results as with a standard evaporation that uses only 
ground-state masses. 
Another problem is that, for hot nuclear fragments in the dense environment, 
the neutron and proton distributions at the surface of 
the fragments can be different from the case of cold isolated nuclei. 
The coefficients $a_{\rm sym}^{\rm vol}$ and $a_{\rm sym}^{\rm surf}$ may change in 
the multifragmentation case. At the present status of the analyses of 
such experiments it is, therefore, reasonable to consider the evolution of only one 
symmetry energy coefficient. 

Of particular interest are the lower values $\gamma \approx 19$~MeV 
obtained for $Z \ge 32$ and 
$^{124}$Sn (Table~I). In a possible interpretation, this observation 
may be related to a nuclear structure effect. Shell effects are properly 
taken into account in the SMM for the ground-state masses which are 
restored at the end of the secondary de-excitation. At earlier stages 
of the de-excitation sequence, however, the liquid-drop description is 
used without shell effects which is adequate at high excitations. 
The decay 
chains ending with the heaviest fragments do not start from a very high 
excitation energy. As shown in Ref.~\cite{buyukcizmeci2013}, a change 
of the symmetry-energy coefficient permits a good description of the 
discontinuities of ground state masses near shell boundaries in a 
liquid-drop description. A switch to $\gamma = 14$~MeV is shown to account 
for the 2-neutron separation energies of nuclei above the neutron 
shell closure and to correspond to the modification of nuclear properties towards 
extreme neutron richness. In the present case, the heaviest product nuclei from 
$^{124}$Sn have neutron numbers $N \ge 50$ while those from $^{112}$Sn 
are below $N=50$. A weak persistence of shell effects may thus be indicated 
for the production of the heaviest fragments. It refers to moderate 
excitations, lower than in typical multifragmentation events. 
The enriched neutron environment may have an additional influence which 
can be expected to grow in importance as one moves closer to the neutron drip-line. 
In particular, short-range correlations, leading to 
proton-neutron pairs, may effectively result in decreasing the 
symmetry energy~\cite{Carb12}.
New fragmentation and multifragmentation experiments with neutron-rich secondary
beams will be important to resolve these questions. 

As discussed previously, multifragmentation reactions can lead to hot nuclei embedded 
in the dense environment of other nuclear species, with all fragments being 
in chemical equilibrium~\cite{Botvina02,LeFevre05,Ogul11}. 
In the present case, the multifragmentation regime covers approximately the production 
of nuclei with atomic number $Z<30$. A similar composition of nuclear matter is also 
typical for certain astrophysical sites at subnuclear densities ($\rho \loo 0.1 \rho_0$) 
as it is, e.g., temporarily realized during the 
collapse and explosion of massive stars~\cite{nihal13}. 

For this reason, the extraction 
of the symmetry energy of hot nuclei from laboratory experiments can 
contribute to the understanding of astrophysical processes which directly depend 
on it, mainly the electron capture and neutrino induced reactions. 
The latter ones, in particular, are crucial for the electron fraction of and the energy 
deposition in star matter which influence the dynamics of supernova explosions. 
As shown in Ref.~\cite{Botvina10}, a reduced symmetry coefficient $\gamma$ affects the density 
dependent electron capture rates on hot nuclei. 
The present analysis confirms the trend of a decreasing symmetry energy as one approaches 
conditions comparable to the multifragmentation regime in agreement with previous 
findings~\cite{Botvina02,LeFevre05,Ogul11,Souliotis07,Hudan09,Ogul09,Buyukcizmeci12}.
The apparent shift and broadening of the isotope mass distributions can be naturally explained 
by a reduced $\gamma$ coefficient. Model studies 
(see, e.g., Refs.~\cite{Carb12,BaLi06,Type10}) suggest that it is more likely the reduced density 
and the influence of the environment, rather than the increased temperature, which are  responsible 
for such a modification of the properties of hot nuclei. This is confirmed by recent relativistic 
Thomas-Fermi calculations which show that a weak decrease of the symmetry-energy coefficient of  
hot nuclei as a function of the temperature is associated with a large increase of the occupied 
volume~\cite{zhang14}. Fully microscopic many-body calculations may be able to clarify this 
evolution.

\section{Conclusions}

It has been shown that the isotopically resolved yield distributions in the range
of atomic numbers $10\leq Z \leq 45$ from the fragmentation of $^{124}$Sn and
$^{112}$Sn projectiles measured with the FRS at 1 GeV/nucleon  
are well reproduced with statistical calculations
in the SMM framework. The good agreement observed with parameters and source 
distributions obtained previously in the analysis of ALADIN fragmentation data for 
similar reactions at 600 MeV/nucleon supports the universal properties of the ensemble of
excited spectator systems produced during the dynamical stage of the reaction.
The previously observed need for a reduction of the symmetry-energy parameter $\gamma$ 
for the description of intermediate-mass fragments from the multifragmentation regime has
been confirmed and found to extend into the regime of heavier product species. 
 
The isotopic distributions of the heavier groups of products from the $^{112}$Sn fragmentation
are centered close to the evaporation-attractor line while
those of heavier products from $^{124}$Sn are located further away. The memory of the 
neutron-richness of the initial projectile system is seen to be preserved for all isotope 
distributions. Their location as a function of the neutron number as well as their widths 
were shown to be mainly sensitive to the symmetry-energy parameter $\gamma$. Besides the general 
trend of a decreasing $\gamma$ with decreasing fragment mass, i.e. with increasing 
violence of the collision, it is found that a slightly reduced $\gamma$ value, with respect 
to $^{112}$Sn, is required for reproducing the mass distributions of the heaviest fragments 
from the $^{124}$Sn decay.
This observation is tentatively interpreted as being caused by a weak persistence of shell 
effects in the production of these fragments at moderate excitation energies. 

More generally, the obtained results demonstrate the feasibility of investigating  
in the laboratory the properties of nuclear species at subnuclear densities and surrounded 
by other species in the freeze-out environment. Experiments 
of this kind will be particularly useful and necessary for nuclei far from the 
stability line. They can be expected to provide us with experimental inputs to the determination
of the nuclear equation of state and of the nuclear compositions and matter properties
in astrophysical environments of extreme isospin. 

\vspace*{0.2cm}

\begin{acknowledgments}
Fruitful discussions with A. Keli\'{c}-Heil are gratefully acknowledged.
This work was supported by TUBITAK (Turkey) with project number 113F058. 
A.S.B. is supported by the GSI Helmholtzzentrum f\"ur 
Schwerionenforschung GmbH and Hessian initiative for scientific and economic excellence 
(LOEWE) through the Helmholtz International Center for FAIR (HIC for FAIR). R.O. thanks GSI
for hospitality and Y\"OK (Y\"uksek\"ogretim Kurulu) for supporting his short visit
to GSI in the summer of 2013.
\end{acknowledgments}

\end{document}